\documentclass[floatfix,aps,prd,twocolumn,superscriptaddress,showpacs]{revtex4}

\usepackage{amsmath}
\usepackage{amsfonts}
\usepackage{amssymb}
\usepackage{bm}
\usepackage{enumerate}
\usepackage[dvips]{color,graphicx}
\usepackage{epsfig}
\usepackage[figuresright]{rotating}

\begin{document}

\title{Measurement of the EMC Effect in the Deuteron}

\newcommand*{\WM}{College of William and Mary, Williamsburg, Virginia 23187, USA}
\newcommand*{\WMindex}{1}
\affiliation{\WM}
\newcommand*{\ANL}{Argonne National Laboratory, Argonne, Illinois 60439}
\newcommand*{\ANLindex}{2}
\affiliation{\ANL}
\newcommand*{\HAMP}{Hampton University, Hampton, Virginia 23668, USA}
\newcommand*{\HAMPindex}{16}
\affiliation{\HAMP}
\newcommand*{\JLAB}{Thomas Jefferson National Accelerator Facility, Newport News, Virginia 23606, USA}
\newcommand*{\JLABindex}{39}
\affiliation{\JLAB}
\newcommand*{\ODU}{Old Dominion University, Norfolk, Virginia 23529, USA}
\newcommand*{\ODUindex}{33}
\affiliation{\ODU}
\newcommand*{\JMU}{James Madison University, Harrisonburg, Virginia 22807, USA}
\newcommand*{\JMUindex}{18}
\affiliation{\JMU}
\newcommand*{\VIRGINIA}{University of Virginia, Charlottesville, Virginia 22901, USA}
\newcommand*{\VIRGINIAindex}{43}
\affiliation{\VIRGINIA}


\author{K.A.~Griffioen} 
\affiliation{\WM}
\author{J.~Arrington} 
\affiliation{\ANL}
\author{M.E.~Christy}
\affiliation{\HAMP}
\author{R.~Ent}
\affiliation{\JLAB}
\author{N.~Kalantarians}
\affiliation{\HAMP}
\author{C.E.~Keppel}
\affiliation{\JLAB}
\author{S.E.~Kuhn} 
\affiliation{\ODU}
\author{W.~Melnitchouk}
\affiliation{\JLAB}
\author{G.~Niculescu} 
\affiliation{\JMU}
\author{I.~Niculescu} 
\affiliation{\JMU}
\author{S.~Tkachenko}
\affiliation{\VIRGINIA}
\author {J.~Zhang} 
\affiliation{\VIRGINIA}

\date{\today}

\begin{abstract}
We have determined the structure function ratio $R^d_{\rm EMC}=F_2^d/(F_2^n+F_2^p)$ from recently published
$F_2^n/F_2^d$ data taken by the BONuS experiment using CLAS at Jefferson Lab.  This ratio deviates 
from unity, with a slope $dR_{\rm EMC}^{d}/dx= -0.10\pm 0.05$ in the range of Bjorken $x$ from 0.35 to 0.7,
for invariant mass $W>1.4$ GeV and $Q^2>1$ GeV$^2$.  The
observed EMC effect for these kinematics is consistent with conventional 
nuclear physics models that include off-shell corrections, 
as well as with empirical analyses that find the EMC effect proportional to the probability of 
short-range nucleon-nucleon correlations.  
\end{abstract}

\maketitle

\section{Introduction}

In the early 1980s the European Muon Collaboration (EMC) discovered that deep-inelastic scattering from atomic nuclei
is not simply the incoherent sum of scattering from the constituent nucleons  \cite{Aubert:1983xm}.  
Their data suggested that quarks  with longitudinal momentum
fraction $x$ in the range 0.35 to 0.7 were suppressed in bound nucleons, and their observations
were quickly confirmed at SLAC \cite{Bodek:1983ec,Bodek:1983qn}.
The deep-inelastic structure function $F_2^A(x)$ for a nucleus with $A$ nucleons
was compared to the equivalent quantity $F_2^d(x)$ for the deuteron, such
that $R_{\rm EMC}^{A}=  (F_2^A/A)/(F_2^d/2)$.  At intermediate $x$, 
$R_{\rm EMC}^{A}$ is less than unity, and this deviation grows with $A$.  
Over the following three decades, subsequent dedicated measurements
\cite{Dasu:1988ru,Ashman:1988bf,Amaudruz:1991cca,Gomez:1993ri,Seely:2009gt} 
confirmed the EMC
effect with ever-increasing precision for a wide range of nuclei.  Drell-Yan data from
Fermilab \cite{Alde:1990im}, however, which were largely sensitive to sea quarks, showed no modifications
of the anti-quark sea for $0.1<x<0.3$, contrary to models predicting anti-quark enhancement.  
Despite many theoretical papers on the EMC effect, no universally accepted explanation has
emerged.  For reviews, see Refs.\ 
\cite{Arneodo:1992wf,Geesaman:1995yd,Norton:2003cb}.  

The precise, new measurements from Jefferson Lab on light nuclei
\cite{Seely:2009gt} have generated a renewed interest in understanding the EMC effect.  
The slopes $|dR_{\rm EMC}^{A}/dx|$ for $0.35<x<0.7$ increase 
with $A$, however, the $^9$Be slope is anomalously large, suggesting perhaps that the EMC 
effect is dependent on local density and that $^9$Be might be acting like two tightly bound 
$\alpha$ particles and a neutron.  A recent analysis \cite{Weinstein:2010rt}  suggests that $dR_{\rm EMC}^{A}/dx$
is proportional to the
probability of finding short-range correlations in nuclei
\cite{Frankfurt:1993sp,Egiyan:2003vg,Fomin:2011ng,Arrington:2011xs,Hen:2014nza,Hen:2012fm}.  Recent work on this subject 
\cite{Hen:2013oha,Atti:2007vx,Melnitchouk:1996vp,Arrington:2012ax,Kulagin:2006dg,Kahn:2008nq} concludes that
although binding and Fermi motion effects contribute, some modification of the bound nucleon's structure
appears to be required to explain the EMC effect. Whether this is caused by the nuclear mean field, short-range correlations, or both
is still  open to debate.  

\begin{figure}[ht]
\includegraphics[width=9.8cm, angle=270]{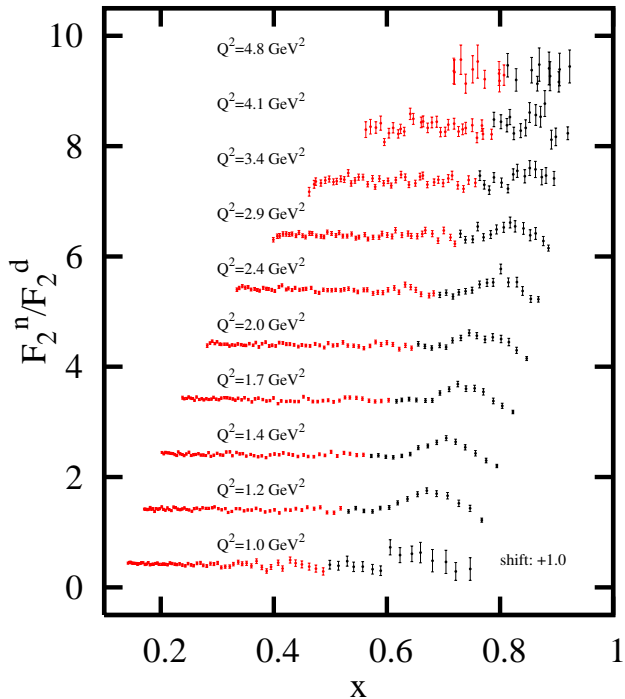}
\caption{(color online). BONuS data for $F_2^n/F_2^d$ vs.~Bjorken $x$ taken with a 5.26 GeV beam.  Only data for
$Q^2\ge 1$ GeV$^2$ are shown.  The red points ($W>1.4$ GeV) are used in this analysis. 
Error bars are statistical only.  Each spectrum is shifted upward by 1.0 from the set below it.
}
\label{fig:5GeV}
\color{black}

\end{figure}

EMC ratios are usually taken with respect to the deuteron, which
is the best proxy for an isoscalar nucleon (neutron plus proton), but the deuteron too
may exhibit an EMC effect.  Several data-driven, model-dependent attempts 
\cite{Lednicky:1990xe,Gomez:1993ri,Weinstein:2010rt} have been made to determine $R_{\rm EMC}^d = F_2^d/(F_2^n+F_2^p)$, 
in which $F_2^{n(p)}$ is the free neutron (proton) structure function.
However, the lack of knowledge about the free neutron's structure has clouded these efforts.  Theoretical
estimates of the deuteron EMC ratio have also been made \cite{West:1972qj, Atwood:1972zp, Frankfurt:1976hb, Kusno:1979dk,
Kaptar:1991hx, Nakano:1991kh, Melnitchouk:1994rv, Braun:1993nh, Burov:1998kz, Kulagin:2004ie,
Arrington:2011qt,Arrington:2008zh,Melnitchouk:1995fc},
often with the goal of isolating $F_2^n/F_2^p$.

A clean measurement of  $R_{\rm EMC}^d$ is greatly needed.  The deuteron is weakly bound
(by 2.2 MeV), and the nucleons are governed only by the $pn$ interaction.  Therefore, a precise measurement
of $R_{\rm EMC}^d$ can shed light on the cause of the EMC effect. 
Because the deuteron has a weak mean field (1 MeV/nucleon binding versus 8 MeV/nucleon for heavier nuclei), but a 
substantial contribution from high-momentum $pn$ pairs, it is a good test-case. 

\section{Data Analysis}

A new extraction of $R_{\rm EMC}^d$ with smaller uncertainties on $F_2^n$ is now possible 
thanks to the high-quality data from the BONuS experiment  
\cite{Fenker:2008zz,Baillie:2011za,Tkachenko:2014byy} using CLAS at Jefferson Lab with electron beams up to 5.26 GeV.
BONuS was designed to measure the high-$x$ 
structure function ratio $F_2^n/F_2^p$ using a model-independent 
extraction of $F_2^n$  that relies on the spectator tagging
technique.  The experiment used a 7-atmosphere gaseous deuterium target surrounded by a 
radial time projection
chamber capable of detecting recoil protons in the range 70-200 MeV/c 
\cite{Fenker:2008zz}.  By selecting backward-going and low-momentum 
spectators, final-state $pn$ interactions and 
off-shell effects were minimized, respectively \cite{Tkachenko:2014byy}.  Detection of the spectator proton ensured 
that the electron scattered from the neutron. The initial-state kinematics of the neutron 
were then calculated from the spectator momentum.  
This technique enabled the extraction of $F_2^n/F_2^d$ over a wide range 
of $x$ for 4-momentum transfer squared $Q^2$ between 0.7 
and 4.5 GeV$^2$,  which covers the resonance region and part of the deep-inelastic region.  
For the present analysis we have used the published data from the 4.22 and 5.26 GeV beam energies
with $Q^2\ge 1$ GeV$^2$ and invariant final-state mass $ W> 1.4$ GeV to determine  $R_{\rm EMC}^d$.

The primary data from BONuS are the ratios $F_2^n/F_2^d$ obtained from measuring tagged
neutron event rates in CLAS and dividing them by the untagged deuteron rates recorded simultaneously 
at the same kinematics \cite{Tkachenko:2014byy}.  Consequently, detector acceptance and other systematic effects 
largely cancel, and the accuracy of this ratio is far better than that of $F_2^n$ alone.

The overall normalization of the BONuS data, which takes into account the spectator proton detection
efficiency, was initially chosen \cite{Baillie:2011za} to make $F_2^n/F_2^p$ at $x=0.3$ agree with the 
CTEQ-Jefferson Lab (CJ) \cite{Accardi:2011fa}
global fit for this point.  There is a 3\% normalization uncertainty
associated with this choice. For the final BONuS results \cite{Tkachenko:2014byy}, which include the resonance region,
the normalization minimized the $\chi^2$ of the full data set with respect to the most recent
update \cite{Kalantarians:2015} of the Christy and Bosted (CB) fits  \cite{Christy:2007ve,Bosted:2007xd}. 
In this case, the convolution model of Ref.~\cite{Kahn:2008nq, Kulagin:2004ie} allowed for a 
self-consistent extraction of $F_2^n$ from $F_2^p$ and $F_2^d$ and better control over the relative normalization of $F_2^n$ and $F_2^d$. 
The new model produced  no change in the 5 GeV normalization, but a 10\% increase in the magnitude of the 4 GeV data. 

Figure~\ref{fig:5GeV} shows the BONuS $F_2^n/F_2^d$ data set taken with a 5.26 GeV beam.
The red points correspond to values of the struck neutron's 
invariant mass $W$ above 1.4 GeV, whereas the black
points ($W<1.4$ GeV) are excluded from this analysis to eliminate the $\Delta$ resonance.

With the new normalization, both the 5.26 and 4.22  GeV data sets yield consistent results within the statistical uncertainties.
To explore the region $x>0.45$ we pushed our analysis into the resonance region ($1.4<W<2.0$ GeV). 
Available data, albeit at slightly higher $Q^2$,  suggest that
$R^A_{\rm EMC}$ in the resonance region is similar to that in the 
deep-inelastic scattering region at the same $x$ \cite{Arrington:2003nt}.
Therefore, we expect that an average over
many different $Q^2$ values washes out any resonance structure and that duality 
ensures $R_{\rm EMC}^d$ at fixed $x$,  averaged over $W$,  approaches the deep-inelastic limit.
These assumptions were tested and confirmed within statistical and systematic uncertainties
by looking for a $Q^2$ dependence of $R^d_{\rm EMC}$ within each $x$-bin
and by considering variations in $R^d_{\rm EMC}$ among four kinematic cases: 
\begin{enumerate}
\item 
$W>1.4$ GeV and $Q^2>1$ GeV$^2$; 
\item
$W>1.8$ GeV and $Q^2>1$ GeV$^2$; 
\item
$W>2.0$ GeV and $Q^2>1$ GeV$^2$; and
\item
$W>2.0$ GeV and $Q^2>2$ GeV$^2$.  
\end{enumerate}
The $F_2^n/F_2^d$ data were sorted into 20-MeV-wide $W$ bins and into logarithmic $Q^2$ bins (13 per decade) with edges at
0.92, 1.10, 1.31, 1.56, 1.87, 2.23, 2,66, 3.17, 3.79, 4.52, and 5.40 GeV$^2$.

The analysis consisted of forming the quantity
\begin{equation}
r(W,Q^2) = \frac{F_2^n}{F_2^d} + \frac{F_2^p}{F_2^d},
\label{eq:r}
\end{equation}
in which the first term is the measured BONuS ratio and the second term is the parameterization of world data
\cite{Christy:2007ve,Bosted:2007xd,Kalantarians:2015}.
All data falling within one of the 20 $x$-bins of width 0.05 were combined using
\begin{eqnarray}
\langle x \rangle &=& \sum_i \frac{x_i}{\sigma_i^2}/ \sum_i \frac{1}{\sigma_i^2},\\
\langle r \rangle  &=& \sum_i \frac{r_i}{\sigma_i^2}/{\sum_i \frac{1}{\sigma_i^2}},\\
\Delta r_{\rm stat} &=& {\sqrt{1/\sum_i \frac{1}{\sigma_i^2}}},\quad{\rm and}\\
\Delta r_{\rm sys}  &=& \sum_i \frac{\Delta r_{{\rm sys},i}}{\sigma_i^2}/{\sum_i \frac{1}{\sigma_i^2}},
\label{eq:comb}
\end{eqnarray}
in which $\sigma_i$ are the statistical uncertainties and $\Delta r_{{\rm sys},i}$ are the corresponding systematic uncertainties
for the $i$th $F_2^n/F_2^d$ datum.


The final values for  $R_{\rm EMC}^d$ were then calculated as
\begin{eqnarray}
R_{\rm EMC}^d &=& {1}/{\langle r\rangle},\\
\Delta R^{\rm stat}_{\rm EMC} &=& {\Delta r_{\rm stat} }/{\langle r\rangle ^2},\quad{\rm and}\\
\Delta R^{\rm sys}_{\rm EMC} &=& {\Delta r_{\rm sys} }/{\langle r\rangle ^2}.
\end{eqnarray}
\section{Uncertainties}
Several checks on our results were made.  First, the analysis was performed by directly calculating $R_{\rm EMC}^d=\left<1/r\right>$
using the same 20 $x$-bins.
The final answers were nearly identical to those in which inversion was the last step.  
The statistical spread in the ratio $r$ in each $x$-bin was used to calculate a
standard error.  This error agreed very well with $\Delta r_{\rm stat} $, which 
supports the hypothesis that
variations in $r$ within a bin are purely statistical.  Systematic bias was also
studied using a cut for $Q^2>2$ GeV$^2$, which in the region of comparison
showed no significant deviation from the data that include lower $Q^2$ values.

Overall systematic uncertainties were estimated by varying the models for $F_2^p/F_2^d$ and the kinematic cuts.
The model dependence was explored using the published CB fits and two later improvements applied to kinematic Case 1 using the 5 GeV data.
The kinematic-dependence was explored using kinematic Cases 1--4 for the 5 GeV data and Case 1 for the 4 GeV data.
In order to separate the overall normalization uncertainty from other systematic uncertainties, we fit the EMC slope in the range $0.35<x<0.7$ and rescaled the data
such that the linear fit intersected unity at $x=0.31$.  This value was obtained from a global analysis of the EMC effect in
all nuclei \cite{Weinstein:2010rt}. The scaling factors ranged from 0.99 to 1.01 for the different cases.  
The average variation in $R_{\rm EMC}^d(x)$ at fixed $x$ for the different cases,  the 1\% scale uncertainty, and
the BONuS systematic uncertainty  $\Delta R_{\rm EMC}^{\rm sys}$ were added in quadrature to yield
 $\Delta R^{\rm sys}_{\rm tot}$, which is listed in Table~\ref{results} and shown as the blue band in Figure~\ref{fig:emc}.
The systematic uncertainties of the BONuS data themselves dominate at large $x$, whereas the model uncertainties of the global fits 
dominate at low $x$ (high $W$).  The mid-$x$ region is dominated by the normalization uncertainty.
For Case 2 with $x>0.4$,  $R^d_{\rm EMC}$ tends to be higher than for Case 1.  This arises in a region of significantly lower statistics
on account of the higher $W$-cut and fewer kinematic points available for resonance averaging.  Although the slope $dR^d_{\rm EMC}/dx$ in this case
is consistent with zero, we find this result unstable to small changes in kinematics.  Case 2 at high $x$ figures into the systematic errors on
our quoted $R^d_{\rm EMC}$ values, however.

Since the data span a large and relatively low $Q^2$  range starting at 1 GeV$^2$, one needs to worry about whether
$R_{\rm EMC}^d$ is simply an artifact of structure function evolution.  To study this we looked at the contents of each
$x$-bin separately.  Figure~\ref{fig:5GeV} shows that each  $x$-bin covers a 
wide enough $Q^2$ range to study $Q^2$ variations within that bin. For this study each data point was
converted into  $R_{\rm EMC}^d$
as described above, and instead of averaging, all values were fit to a straight line vs.\ $Q^2$.  
Fitting to a constant slope yields $dR_{\rm EMC}^d/dQ^2=0.0037(45)$, which is consistent with no observable $Q^2$ variation.

Although the BONuS $F_2$ data were extracted assuming that the longitudinal-to-transverse cross 
section ratio $R$ cancels in the neutron to deuteron ratios, the associated uncertainty is included in
the published results.  Some nuclear dependence to $R$ could, however, slightly modify our EMC results
\cite{Guzey:2012yk}.


\section{Results}

\begin{figure}[ht]
\includegraphics[width=6.3cm, angle=270]{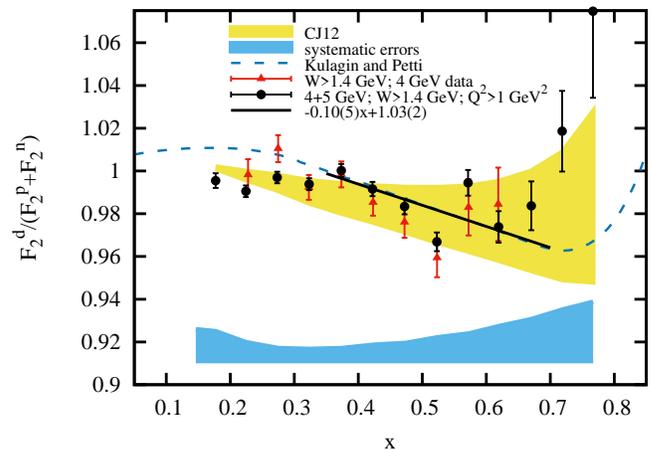}
\caption{(color online)
The deuteron EMC ratio $R_{\rm EMC}^d= F_2^d/(F_2^n+F_2^p)$ as extracted from the BONuS data.
Total systematic uncertainties are shown as a band arbitrarily positioned
at 0.91 (blue). The yellow band shows the CJ12 \protect\cite{Owens:2012bv} limits expected from their nuclear models.
The black points are the combined 4 and 5 GeV data, whereas the red points are the 4 GeV data alone.
The dashed blue line shows the calculations of Ref.~\protect\cite{Kulagin:2004ie}.
The solid line (black) is the fit to the black points for $0.35<x<0.7$.
}
\color{black}
\label{fig:emc}
\end{figure}

Our final result uses the new self-consistent convolution model \cite{Kalantarians:2015} for $F_2^p/F_2^d$, which 
was used to determine the absolute normalization of the final published BONuS $F_2^n/F_2^d$ data \cite{Tkachenko:2014byy}. 
It provides an excellent
representation of $F_2$ for our kinematics.  Our result uses the combined 5.26 and 4.22 data with
cuts $Q^2>1$ GeV$^2$ and $W>1.4$ GeV.
A linear fit for $0.35<x<0.7$ yields $dR_{\rm EMC}^d/dx=-0.10 \pm 0.05$ 
where the uncertainty comes from the $\chi^2$ fit.  
Figure~\ref{fig:emc} shows these results together with comparisons to various models.
For $x<0.5$ the EMC ratios  $R_{\rm EMC}^d$ agree within uncertainties with those obtained using
more stringent cuts in $W$.  The ratio for $x>0.5$ continues the trend
of the lower-$x$ data, with a hint of the expected rise above $x=0.7$ as seen in  $R_{\rm EMC}^{A}$ for heavier nuclei,
but these high-$x$ values are more uncertain because there are fewer data points for resonance averaging.
The black circles are the combined results for 4 and 5 GeV,  which are clearly dominated by the 5 GeV data. 
The 4 GeV data by themselves (red triangles), are consistent with the combined data set.  
The two points between $x=0.5$ and 0.6 seem to be off the trend,
one being high and the other low.  Because this is consistent for the two beam energies, we suspect that 
there is a slight mismatch between the model form factors and the data in this region.

Table~\ref{results} gives our numerical results, in which $N$ is the number of $F_2^n/F_2^d$ points contributing to a bin with
average kinematic values $\langle x\rangle$ and $\langle Q^2\rangle$.  Here $\Delta R_{\rm EMC}^{\rm stat}$ and $\Delta R_{\rm EMC}^{\rm sys}$
are the statistical and systematic uncertainties that come from the BONuS data themselves, and  $\Delta R^{\rm sys}_{\rm tot}$ is
the total systematic uncertainty that includes $\Delta R_{\rm EMC}^{\rm sys}$ plus the modeling and normalization uncertainties
in $F_2^p/F_2^d$.

\begin {table} [ht]
\begin{center}
\caption {EMC results for the deuteron.
The columns correspond to the number of kinematic points, average $x$ and $Q^2$, the EMC ratio,
the statistical and systematic errors from the BONuS data, and the total systematic error including 
modeling of $F_2^p/F_2^d$.}

\vskip10pt
\begin{tabular} {|c|c|c|c|c|c|c|} \hline
$N$ & $\langle x\rangle$ & $\langle Q^2\rangle$ & $R_{\rm EMC}^d$ & $\Delta R_{\rm EMC}^{\rm stat}$ & 
$\Delta R_{\rm EMC}^{\rm sys}$ & $\Delta R^{\rm sys}_{\rm tot}$ \\ 
&&(GeV$^2)$  & & & &\\
\hline
~28~ & ~0.177~ &  1.09 &  ~0.995~ & ~0.003~ & ~0.002~   & ~0.015~ \\
 55 & 0.224 &  1.24 &  0.991 & 0.003 & 0.003   & 0.010 \\
 65 & 0.273 &  1.39 &  0.997 & 0.003 & 0.003   & 0.007 \\
 71 & 0.323 &  1.50 &  0.994 & 0.003 & 0.004   & 0.007 \\
 70 & 0.373 &  1.63 &  1.000 & 0.003 & 0.005   & 0.007 \\
 70 & 0.422 &  1.71 &  0.992 & 0.003 & 0.007   & 0.009 \\
 71 & 0.472 &  1.85 &  0.983 & 0.004 & 0.009   & 0.009 \\
 56 & 0.523 &  2.01 &  0.967 & 0.004 & 0.011   & 0.012 \\
 47 & 0.572 &  2.30 &  0.994 & 0.006 & 0.013   & 0.014 \\
 41 & 0.619 &  2.54 &  0.974 & 0.007 & 0.017   & 0.017 \\
 26 & 0.670 &  2.97 &  0.984 & 0.011 & 0.020   & 0.021 \\
 21 & 0.719 &  3.39 &  1.019 & 0.019 & 0.023   & 0.025 \\
 11 & 0.767 &  4.03 &  1.075 & 0.041 & 0.024   & 0.029 \\
\hline
\end{tabular}
\label{results}
\end{center}
\end{table}

The current results can be compared to the SLAC model-dependent extraction from Ref.~\cite{Gomez:1993ri}.
Here $R_{\rm EMC}^d$ was estimated assuming the hypothesis of Ref.~\cite{Frankfurt:1988nt} that 
$1+R_{\rm EMC}$ is proportional to the nucleon density.  
The SLAC slope $dR_{\rm EMC}^d/dx = -0.098\pm 0.005$ is similar to our own, but its quoted uncertainty 
takes no account of the model-dependence.
The assumption of density-dependence gives consistent results with
our measurements for the deuteron.  Semi-empirical models like that of Ref.~\cite{Kulagin:2004ie} (blue dashed
curve in Figure~\ref{fig:emc}), which include Fermi motion, binding,
and off-shell effects, are able to describe the shape of $R^d_{\rm EMC}$ quite well.  
Our data are also consistent with the CJ12 \cite{Owens:2012bv} band in yellow. 

We have explored whether the Nachtmann variable $\xi = 2x/(1+\sqrt{1+4M^2x^2/Q^2})$ (with $M$ the nucleon mass)
would be better suited than $x$ to represent $R^d_{\rm EMC}$, since our data are at relatively 
low $Q^2$.  The authors of Refs.~\cite{Arrington:2003nt,Seely:2009gt} too have addressed this question. They and we prefer $x$, which has been the
common variable of discourse and calculation.  
Our EMC ratios are determined using data and model at precisely the same values of $W$ and $Q^2$.  Therefore, plotting versus $\xi$ merely redistributes
the EMC points along the $x$ axis.  Generally, $\xi$ is smaller than $x$.  Consequently, more of the high-$x$ resonances in the data-set now contribute to the EMC
slope.  Thus, using $\xi$ to reduce the effect of resonances, actually increases their influence.  
A fit over the rescaled interval [0.35,0.65] yields $dR^d_{\rm EMC}/d\xi = -0.08\pm 0.06$.  The slope is slightly smaller and
the uncertainty slightly larger than when we plot versus $x$. Resonance states above $x=0.7$ drive the slope to slightly smaller values than the fit versus $x$.

The analysis of Ref.~\cite{Weinstein:2010rt} finds a linear relationship of
the EMC slopes $dR_{\rm EMC}^A/dx$ versus the relative
short-range correlation probability  $R_{2N}(A/d)$ in a nucleus $A$ with respect to the deuteron. From
that analysis the authors conclude that the deuteron EMC slope should be  $dR_{\rm EMC}^d/dx =-0.079\pm 0.006$.  This
value is somewhat smaller than our result of $-0.10\pm 0.05$ but is consistent within 1$\sigma$.  
A more recent analysis along these same lines brackets the slope between $-0.079$ and $-0.106$ \cite{Hen:2012fm}, 
and suggests that
the uncertainties of Ref.~\cite{Weinstein:2010rt} are underestimated.  

\section{$R^d_{\rm EMC}$ and Short-Range Correlations}

\begin{figure}[ht]
\includegraphics[width=6.3cm, angle=270]{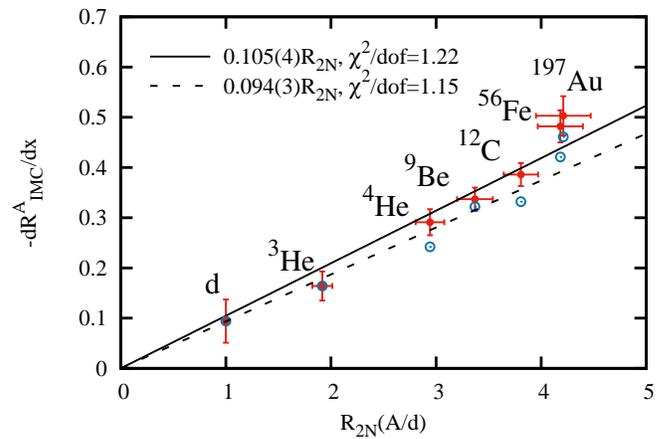}
\caption{(color online)
EMC slopes per isoscalar nucleon, $-dR_{\rm IMC}^{A}/dx$, versus the relative probability with respect to the
deuteron of short-range correlations, $R_{2N}(A/d)$.
Fits assume that ($dR_{\rm IMC}^{A}/dx)/(R_{2N}(A/d)$) is constant. The red points
are from Ref.~\protect\cite{Weinstein:2010rt}. The blue points
are from  Ref.~\protect\cite{Hen:2013oha} and are corrected for isospin and for $x$ normalized to a maximum of $x=A$.
Their uncertainties are the same as for the red points.
}
\color{black}
\label{fig:weinstein}
\end{figure}

We are able
to use our results to estimate the in-medium correction 
$R^A_{\rm IMC} = 2F_2^A/A(F_2^n+F_2^p)$ with slope $dR_{\rm IMC}^A/dx$, 
for which the normalizing factor is the isoscalar free nucleon.   
We write $R_{\rm EMC}^A = 1 + (dR_{\rm EMC}^A/dx)(x_0-x)$ assuming that all nuclei have ratios
of unity at $x_0=0.31\pm 0.04$, as found in Ref.~\cite{Weinstein:2010rt}.  The nuclear EMC ratio
$R_{\rm EMC}^{A}$ can
be multiplied by the deuteron EMC ratio $R_{\rm EMC}^d$ 
to obtain $R_{\rm IMC}^A$.  Hence, to good approximation,
$dR_{\rm IMC}^A/dx =  dR_{\rm EMC}^{A}/dx + dR_{\rm EMC}^d/dx$.  Figure~\ref{fig:weinstein}
shows the results.  The data are consistent with the {\it ansatz} that 
$dR_{\rm IMC}^A/dx$ is directly proportional to
$R_{2N}(A/d)$, the short-range correlation probability, with a proportionality constant 
$0.105\pm 0.004$ ($\chi^2$/dof = 1.22). This effect persists for
the isospin and nuclear-$x$-corrected data from Ref.~\cite{Hen:2013oha} (blue points), which have the same uncertainties as
the red points. 
The linear relationship between short-range correlations and EMC slopes, with the shift for the deuteron EMC effect,
is now consistent with an intercept of zero, and the relationship becomes a straight proportion 
described by a single free parameter.

\section{Synopsis}

In summary, we find an EMC-like slope in the ratio of deuteron to free nucleon structure functions, 
using the BONuS data (which are partially in the nucleon resonance region above the $\Delta$ resonance). 
This slope is consistent with conventional nuclear physics models that include off-shell corrections, as well as
with short-range-correlation models of the EMC effect. 
This first, direct measurement of the magnitude
of the EMC effect in deuterium 
demonstrates that the new BONuS experiment at 11 GeV using 
CLAS12, with its better precision, larger average 
$Q^2$, and deep-inelastic kinematics, will be able to 
determine $R^d_{\rm EMC}$ with good accuracy.

\vspace{-24pt}

\begin{acknowledgments}

\vspace{-10pt}
We thank the staff of the Jefferson Lab accelerator and Hall~B
for their support on the BONuS experiment.
This work was supported by the United States Department of Energy (DOE) Contract No. DE-AC05- 06OR23177, 
under which Jefferson Science Associates, LLC operates Jefferson Lab. 
S.K., J.A., S.T., and K.G.\  
acknowledge support from the DOE, Office of Science, Office of Nuclear Physics, under grants
DE-FG02-96ER40960, DE-AC02-06CH11357, DE-FG02-97ER41025, and DE-FG02-96ER41003, respectively.
I.N.\ and G.N.\ acknowledge support from the NSF under grant  PHY-1307196. M.E.C.\ acknowledges
support from NSF grants PHY-1002644 and PHY-1307415.
\end{acknowledgments}

\bibliography{EMC}

\begin{thebibliography}{50}
\expandafter\ifx\csname natexlab\endcsname\relax\def\natexlab#1{#1}\fi
\expandafter\ifx\csname bibnamefont\endcsname\relax
  \def\bibnamefont#1{#1}\fi
\expandafter\ifx\csname bibfnamefont\endcsname\relax
  \def\bibfnamefont#1{#1}\fi
\expandafter\ifx\csname citenamefont\endcsname\relax
  \def\citenamefont#1{#1}\fi
\expandafter\ifx\csname url\endcsname\relax
  \def\url#1{\texttt{#1}}\fi
\expandafter\ifx\csname urlprefix\endcsname\relax\def\urlprefix{URL }\fi
\providecommand{\bibinfo}[2]{#2}
\providecommand{\eprint}[2][]{\url{#2}}

\bibitem[{\citenamefont{Aubert et~al.}(1983)}]{Aubert:1983xm}
\bibinfo{author}{\bibfnamefont{J.}~\bibnamefont{Aubert}} \bibnamefont{et~al.}
  (\bibinfo{collaboration}{European Muon Collaboration}),
  \bibinfo{journal}{Phys. Lett.} \textbf{\bibinfo{volume}{B123}},
  \bibinfo{pages}{275} (\bibinfo{year}{1983}).

\bibitem[{\citenamefont{Bodek et~al.}(1983{\natexlab{a}})\citenamefont{Bodek,
  Giokaris, Atwood, Coward, Dubin et~al.}}]{Bodek:1983ec}
\bibinfo{author}{\bibfnamefont{A.}~\bibnamefont{Bodek}},
  \bibinfo{author}{\bibfnamefont{N.}~\bibnamefont{Giokaris}},
  \bibinfo{author}{\bibfnamefont{W.}~\bibnamefont{Atwood}},
  \bibinfo{author}{\bibfnamefont{D.}~\bibnamefont{Coward}},
  \bibinfo{author}{\bibfnamefont{D.}~\bibnamefont{Dubin}},
  \bibnamefont{et~al.}, \bibinfo{journal}{Phys. Rev. Lett.}
  \textbf{\bibinfo{volume}{51}}, \bibinfo{pages}{534}
  (\bibinfo{year}{1983}{\natexlab{a}}).

\bibitem[{\citenamefont{Bodek et~al.}(1983{\natexlab{b}})\citenamefont{Bodek,
  Giokaris, Atwood, Coward, Sherden et~al.}}]{Bodek:1983qn}
\bibinfo{author}{\bibfnamefont{A.}~\bibnamefont{Bodek}},
  \bibinfo{author}{\bibfnamefont{N.}~\bibnamefont{Giokaris}},
  \bibinfo{author}{\bibfnamefont{W.}~\bibnamefont{Atwood}},
  \bibinfo{author}{\bibfnamefont{D.}~\bibnamefont{Coward}},
  \bibinfo{author}{\bibfnamefont{D.}~\bibnamefont{Sherden}},
  \bibnamefont{et~al.}, \bibinfo{journal}{Phys. Rev. Lett.}
  \textbf{\bibinfo{volume}{50}}, \bibinfo{pages}{1431}
  (\bibinfo{year}{1983}{\natexlab{b}}).

\bibitem[{\citenamefont{Dasu et~al.}(1988)\citenamefont{Dasu, de~Barbaro,
  Bodek, Harada, Krasny et~al.}}]{Dasu:1988ru}
\bibinfo{author}{\bibfnamefont{S.}~\bibnamefont{Dasu}},
  \bibinfo{author}{\bibfnamefont{P.}~\bibnamefont{de~Barbaro}},
  \bibinfo{author}{\bibfnamefont{A.}~\bibnamefont{Bodek}},
  \bibinfo{author}{\bibfnamefont{H.}~\bibnamefont{Harada}},
  \bibinfo{author}{\bibfnamefont{M.}~\bibnamefont{Krasny}},
  \bibnamefont{et~al.}, \bibinfo{journal}{Phys. Rev. Lett.}
  \textbf{\bibinfo{volume}{60}}, \bibinfo{pages}{2591} (\bibinfo{year}{1988}).

\bibitem[{\citenamefont{Ashman et~al.}(1988)}]{Ashman:1988bf}
\bibinfo{author}{\bibfnamefont{J.}~\bibnamefont{Ashman}} \bibnamefont{et~al.}
  (\bibinfo{collaboration}{European Muon Collaboration}),
  \bibinfo{journal}{Phys. Lett.} \textbf{\bibinfo{volume}{B202}},
  \bibinfo{pages}{603} (\bibinfo{year}{1988}).

\bibitem[{\citenamefont{Amaudruz et~al.}(1991)}]{Amaudruz:1991cca}
\bibinfo{author}{\bibfnamefont{P.}~\bibnamefont{Amaudruz}} \bibnamefont{et~al.}
  (\bibinfo{collaboration}{New Muon Collaboration}), \bibinfo{journal}{Z.
  Phys.} \textbf{\bibinfo{volume}{C51}}, \bibinfo{pages}{387}
  (\bibinfo{year}{1991}).

\bibitem[{\citenamefont{Gomez et~al.}(1994)\citenamefont{Gomez, Arnold, Bosted,
  Chang, Katramatou et~al.}}]{Gomez:1993ri}
\bibinfo{author}{\bibfnamefont{J.}~\bibnamefont{Gomez}},
  \bibinfo{author}{\bibfnamefont{R.}~\bibnamefont{Arnold}},
  \bibinfo{author}{\bibfnamefont{P.~E.} \bibnamefont{Bosted}},
  \bibinfo{author}{\bibfnamefont{C.}~\bibnamefont{Chang}},
  \bibinfo{author}{\bibfnamefont{A.}~\bibnamefont{Katramatou}},
  \bibnamefont{et~al.}, \bibinfo{journal}{Phys. Rev.}
  \textbf{\bibinfo{volume}{D49}}, \bibinfo{pages}{4348} (\bibinfo{year}{1994}).

\bibitem[{\citenamefont{Seely et~al.}(2009)\citenamefont{Seely, Daniel,
  Gaskell, Arrington, Fomin et~al.}}]{Seely:2009gt}
\bibinfo{author}{\bibfnamefont{J.}~\bibnamefont{Seely}},
  \bibinfo{author}{\bibfnamefont{A.}~\bibnamefont{Daniel}},
  \bibinfo{author}{\bibfnamefont{D.}~\bibnamefont{Gaskell}},
  \bibinfo{author}{\bibfnamefont{J.}~\bibnamefont{Arrington}},
  \bibinfo{author}{\bibfnamefont{N.}~\bibnamefont{Fomin}},
  \bibnamefont{et~al.}, \bibinfo{journal}{Phys. Rev. Lett.}
  \textbf{\bibinfo{volume}{103}}, \bibinfo{pages}{202301}
  (\bibinfo{year}{2009}), \eprint{0904.4448}.

\bibitem[{\citenamefont{Alde et~al.}(1990)\citenamefont{Alde, Baer, Carey,
  Garvey, Klein et~al.}}]{Alde:1990im}
\bibinfo{author}{\bibfnamefont{D.}~\bibnamefont{Alde}},
  \bibinfo{author}{\bibfnamefont{H.}~\bibnamefont{Baer}},
  \bibinfo{author}{\bibfnamefont{T.}~\bibnamefont{Carey}},
  \bibinfo{author}{\bibfnamefont{G.}~\bibnamefont{Garvey}},
  \bibinfo{author}{\bibfnamefont{A.}~\bibnamefont{Klein}},
  \bibnamefont{et~al.}, \bibinfo{journal}{Phys. Rev. Lett.}
  \textbf{\bibinfo{volume}{64}}, \bibinfo{pages}{2479} (\bibinfo{year}{1990}).

\bibitem[{\citenamefont{Arneodo}(1994)}]{Arneodo:1992wf}
\bibinfo{author}{\bibfnamefont{M.}~\bibnamefont{Arneodo}},
  \bibinfo{journal}{Phys. Rept.} \textbf{\bibinfo{volume}{240}},
  \bibinfo{pages}{301} (\bibinfo{year}{1994}).

\bibitem[{\citenamefont{Geesaman et~al.}(1995)\citenamefont{Geesaman, Saito,
  and Thomas}}]{Geesaman:1995yd}
\bibinfo{author}{\bibfnamefont{D.~F.} \bibnamefont{Geesaman}},
  \bibinfo{author}{\bibfnamefont{K.}~\bibnamefont{Saito}}, \bibnamefont{and}
  \bibinfo{author}{\bibfnamefont{A.~W.} \bibnamefont{Thomas}},
  \bibinfo{journal}{Ann. Rev. Nucl. Part. Sci.} \textbf{\bibinfo{volume}{45}},
  \bibinfo{pages}{337} (\bibinfo{year}{1995}).

\bibitem[{\citenamefont{Norton}(2003)}]{Norton:2003cb}
\bibinfo{author}{\bibfnamefont{P.}~\bibnamefont{Norton}},
  \bibinfo{journal}{Rept. Prog. Phys.} \textbf{\bibinfo{volume}{66}},
  \bibinfo{pages}{1253} (\bibinfo{year}{2003}).

\bibitem[{\citenamefont{Weinstein et~al.}(2011)\citenamefont{Weinstein,
  Piasetzky, Higinbotham, Gomez, Hen et~al.}}]{Weinstein:2010rt}
\bibinfo{author}{\bibfnamefont{L.}~\bibnamefont{Weinstein}},
  \bibinfo{author}{\bibfnamefont{E.}~\bibnamefont{Piasetzky}},
  \bibinfo{author}{\bibfnamefont{D.}~\bibnamefont{Higinbotham}},
  \bibinfo{author}{\bibfnamefont{J.}~\bibnamefont{Gomez}},
  \bibinfo{author}{\bibfnamefont{O.}~\bibnamefont{Hen}}, \bibnamefont{et~al.},
  \bibinfo{journal}{Phys. Rev. Lett.} \textbf{\bibinfo{volume}{106}},
  \bibinfo{pages}{052301} (\bibinfo{year}{2011}), \eprint{1009.5666}.

\bibitem[{\citenamefont{Frankfurt et~al.}(1993)\citenamefont{Frankfurt,
  Strikman, Day, and Sargsian}}]{Frankfurt:1993sp}
\bibinfo{author}{\bibfnamefont{L.}~\bibnamefont{Frankfurt}},
  \bibinfo{author}{\bibfnamefont{M.}~\bibnamefont{Strikman}},
  \bibinfo{author}{\bibfnamefont{D.}~\bibnamefont{Day}}, \bibnamefont{and}
  \bibinfo{author}{\bibfnamefont{M.}~\bibnamefont{Sargsian}},
  \bibinfo{journal}{Phys. Rev.} \textbf{\bibinfo{volume}{C48}},
  \bibinfo{pages}{2451} (\bibinfo{year}{1993}).

\bibitem[{\citenamefont{Egiyan et~al.}(2003)}]{Egiyan:2003vg}
\bibinfo{author}{\bibfnamefont{K.}~\bibnamefont{Egiyan}} \bibnamefont{et~al.}
  (\bibinfo{collaboration}{CLAS}), \bibinfo{journal}{Phys.Rev.}
  \textbf{\bibinfo{volume}{C68}}, \bibinfo{pages}{014313}
  (\bibinfo{year}{2003}), \eprint{nucl-ex/0301008}.

\bibitem[{\citenamefont{Fomin et~al.}(2012)\citenamefont{Fomin, Arrington,
  Asaturyan, Benmokhtar, Boeglin et~al.}}]{Fomin:2011ng}
\bibinfo{author}{\bibfnamefont{N.}~\bibnamefont{Fomin}},
  \bibinfo{author}{\bibfnamefont{J.}~\bibnamefont{Arrington}},
  \bibinfo{author}{\bibfnamefont{R.}~\bibnamefont{Asaturyan}},
  \bibinfo{author}{\bibfnamefont{F.}~\bibnamefont{Benmokhtar}},
  \bibinfo{author}{\bibfnamefont{W.}~\bibnamefont{Boeglin}},
  \bibnamefont{et~al.}, \bibinfo{journal}{Phys. Rev. Lett.}
  \textbf{\bibinfo{volume}{108}}, \bibinfo{pages}{092502}
  (\bibinfo{year}{2012}), \eprint{1107.3583}.

\bibitem[{\citenamefont{Arrington
  et~al.}(2012{\natexlab{a}})\citenamefont{Arrington, Higinbotham, Rosner, and
  Sargsian}}]{Arrington:2011xs}
\bibinfo{author}{\bibfnamefont{J.}~\bibnamefont{Arrington}},
  \bibinfo{author}{\bibfnamefont{D.}~\bibnamefont{Higinbotham}},
  \bibinfo{author}{\bibfnamefont{G.}~\bibnamefont{Rosner}}, \bibnamefont{and}
  \bibinfo{author}{\bibfnamefont{M.}~\bibnamefont{Sargsian}},
  \bibinfo{journal}{Prog. Part. Nucl. Phys.} \textbf{\bibinfo{volume}{67}},
  \bibinfo{pages}{898} (\bibinfo{year}{2012}{\natexlab{a}}),
  \eprint{1104.1196}.

\bibitem[{\citenamefont{Hen et~al.}(2014)\citenamefont{Hen, Sargsian,
  Weinstein, Piasetzky, Hakobyan et~al.}}]{Hen:2014nza}
\bibinfo{author}{\bibfnamefont{O.}~\bibnamefont{Hen}},
  \bibinfo{author}{\bibfnamefont{M.}~\bibnamefont{Sargsian}},
  \bibinfo{author}{\bibfnamefont{L.}~\bibnamefont{Weinstein}},
  \bibinfo{author}{\bibfnamefont{E.}~\bibnamefont{Piasetzky}},
  \bibinfo{author}{\bibfnamefont{H.}~\bibnamefont{Hakobyan}},
  \bibnamefont{et~al.}, \bibinfo{journal}{Science}
  \textbf{\bibinfo{volume}{346}}, \bibinfo{pages}{614} (\bibinfo{year}{2014}),
  \eprint{1412.0138}.

\bibitem[{\citenamefont{Hen et~al.}(2012)\citenamefont{Hen, Piasetzky, and
  Weinstein}}]{Hen:2012fm}
\bibinfo{author}{\bibfnamefont{O.}~\bibnamefont{Hen}},
  \bibinfo{author}{\bibfnamefont{E.}~\bibnamefont{Piasetzky}},
  \bibnamefont{and}
  \bibinfo{author}{\bibfnamefont{L.}~\bibnamefont{Weinstein}},
  \bibinfo{journal}{Phys. Rev.} \textbf{\bibinfo{volume}{C85}},
  \bibinfo{pages}{047301} (\bibinfo{year}{2012}), \eprint{1202.3452}.

\bibitem[{\citenamefont{Hen et~al.}(2013)\citenamefont{Hen, Higinbotham,
  Miller, Piasetzky, and Weinstein}}]{Hen:2013oha}
\bibinfo{author}{\bibfnamefont{O.}~\bibnamefont{Hen}},
  \bibinfo{author}{\bibfnamefont{D.}~\bibnamefont{Higinbotham}},
  \bibinfo{author}{\bibfnamefont{G.}~\bibnamefont{Miller}},
  \bibinfo{author}{\bibfnamefont{E.}~\bibnamefont{Piasetzky}},
  \bibnamefont{and}
  \bibinfo{author}{\bibfnamefont{L.}~\bibnamefont{Weinstein}},
  \bibinfo{journal}{Int. J. Mod. Phys.} \textbf{\bibinfo{volume}{E22}},
  \bibinfo{pages}{1330017} (\bibinfo{year}{2013}), \eprint{1304.2813}.

\bibitem[{\citenamefont{Ciofi~degli Atti et~al.}(2007)\citenamefont{Ciofi~degli
  Atti, Frankfurt, Kaptari, and Strikman}}]{Atti:2007vx}
\bibinfo{author}{\bibfnamefont{C.}~\bibnamefont{Ciofi~degli Atti}},
  \bibinfo{author}{\bibfnamefont{L.}~\bibnamefont{Frankfurt}},
  \bibinfo{author}{\bibfnamefont{L.}~\bibnamefont{Kaptari}}, \bibnamefont{and}
  \bibinfo{author}{\bibfnamefont{M.}~\bibnamefont{Strikman}},
  \bibinfo{journal}{Phys. Rev.} \textbf{\bibinfo{volume}{C76}},
  \bibinfo{pages}{055206} (\bibinfo{year}{2007}), \eprint{0706.2937}.

\bibitem[{\citenamefont{Melnitchouk et~al.}(1997)\citenamefont{Melnitchouk,
  Sargsian, and Strikman}}]{Melnitchouk:1996vp}
\bibinfo{author}{\bibfnamefont{W.}~\bibnamefont{Melnitchouk}},
  \bibinfo{author}{\bibfnamefont{M.}~\bibnamefont{Sargsian}}, \bibnamefont{and}
  \bibinfo{author}{\bibfnamefont{M.}~\bibnamefont{Strikman}},
  \bibinfo{journal}{Z. Phys.} \textbf{\bibinfo{volume}{A359}},
  \bibinfo{pages}{99} (\bibinfo{year}{1997}), \eprint{nucl-th/9609048}.

\bibitem[{\citenamefont{Arrington
  et~al.}(2012{\natexlab{b}})\citenamefont{Arrington, Daniel, Day, Fomin,
  Gaskell et~al.}}]{Arrington:2012ax}
\bibinfo{author}{\bibfnamefont{J.}~\bibnamefont{Arrington}},
  \bibinfo{author}{\bibfnamefont{A.}~\bibnamefont{Daniel}},
  \bibinfo{author}{\bibfnamefont{D.}~\bibnamefont{Day}},
  \bibinfo{author}{\bibfnamefont{N.}~\bibnamefont{Fomin}},
  \bibinfo{author}{\bibfnamefont{D.}~\bibnamefont{Gaskell}},
  \bibnamefont{et~al.}, \bibinfo{journal}{Phys. Rev.}
  \textbf{\bibinfo{volume}{C86}}, \bibinfo{pages}{065204}
  (\bibinfo{year}{2012}{\natexlab{b}}), \eprint{1206.6343}.

\bibitem[{\citenamefont{Kulagin and
  Petti}(2006{\natexlab{a}})}]{Kulagin:2006dg}
\bibinfo{author}{\bibfnamefont{S.~A.} \bibnamefont{Kulagin}} \bibnamefont{and}
  \bibinfo{author}{\bibfnamefont{R.}~\bibnamefont{Petti}},
  \bibinfo{journal}{Nucl. Phys. Proc. Suppl.} \textbf{\bibinfo{volume}{159}},
  \bibinfo{pages}{180} (\bibinfo{year}{2006}{\natexlab{a}}),
  \eprint{hep-ph/0602090}.

\bibitem[{\citenamefont{Kahn et~al.}(2009)\citenamefont{Kahn, Melnitchouk, and
  Kulagin}}]{Kahn:2008nq}
\bibinfo{author}{\bibfnamefont{Y.}~\bibnamefont{Kahn}},
  \bibinfo{author}{\bibfnamefont{W.}~\bibnamefont{Melnitchouk}},
  \bibnamefont{and} \bibinfo{author}{\bibfnamefont{S.~A.}
  \bibnamefont{Kulagin}}, \bibinfo{journal}{Phys. Rev.}
  \textbf{\bibinfo{volume}{C79}}, \bibinfo{pages}{035205}
  (\bibinfo{year}{2009}), \eprint{0809.4308}.

\bibitem[{\citenamefont{Lednicky et~al.}(1990)\citenamefont{Lednicky,
  Peshekhonov, and Smirnov}}]{Lednicky:1990xe}
\bibinfo{author}{\bibfnamefont{R.}~\bibnamefont{Lednicky}},
  \bibinfo{author}{\bibfnamefont{D.}~\bibnamefont{Peshekhonov}},
  \bibnamefont{and} \bibinfo{author}{\bibfnamefont{G.}~\bibnamefont{Smirnov}},
  \bibinfo{journal}{Sov. J. Nucl. Phys.} \textbf{\bibinfo{volume}{52}},
  \bibinfo{pages}{552} (\bibinfo{year}{1990}).

\bibitem[{\citenamefont{West}(1971)}]{West:1972qj}
\bibinfo{author}{\bibfnamefont{G.}~\bibnamefont{West}}, \bibinfo{journal}{Phys.
  Lett.} \textbf{\bibinfo{volume}{B37}}, \bibinfo{pages}{509}
  (\bibinfo{year}{1971}).

\bibitem[{\citenamefont{Atwood and West}(1973)}]{Atwood:1972zp}
\bibinfo{author}{\bibfnamefont{W.}~\bibnamefont{Atwood}} \bibnamefont{and}
  \bibinfo{author}{\bibfnamefont{G.~B.} \bibnamefont{West}},
  \bibinfo{journal}{Phys. Rev.} \textbf{\bibinfo{volume}{D7}},
  \bibinfo{pages}{773} (\bibinfo{year}{1973}).

\bibitem[{\citenamefont{Frankfurt and Strikman}(1978)}]{Frankfurt:1976hb}
\bibinfo{author}{\bibfnamefont{L.}~\bibnamefont{Frankfurt}} \bibnamefont{and}
  \bibinfo{author}{\bibfnamefont{M.}~\bibnamefont{Strikman}},
  \bibinfo{journal}{Phys. Lett.} \textbf{\bibinfo{volume}{B76}},
  \bibinfo{pages}{333} (\bibinfo{year}{1978}).

\bibitem[{\citenamefont{Kusno and Moravcsik}(1981)}]{Kusno:1979dk}
\bibinfo{author}{\bibfnamefont{D.}~\bibnamefont{Kusno}} \bibnamefont{and}
  \bibinfo{author}{\bibfnamefont{M.~J.} \bibnamefont{Moravcsik}},
  \bibinfo{journal}{Nucl. Phys.} \textbf{\bibinfo{volume}{B184}},
  \bibinfo{pages}{283} (\bibinfo{year}{1981}).

\bibitem[{\citenamefont{Kaptar and Umnikov}(1991)}]{Kaptar:1991hx}
\bibinfo{author}{\bibfnamefont{L.}~\bibnamefont{Kaptar}} \bibnamefont{and}
  \bibinfo{author}{\bibfnamefont{A.~Y.} \bibnamefont{Umnikov}},
  \bibinfo{journal}{Phys. Lett.} \textbf{\bibinfo{volume}{B259}},
  \bibinfo{pages}{155} (\bibinfo{year}{1991}).

\bibitem[{\citenamefont{Nakano and Wong}(1991)}]{Nakano:1991kh}
\bibinfo{author}{\bibfnamefont{K.}~\bibnamefont{Nakano}} \bibnamefont{and}
  \bibinfo{author}{\bibfnamefont{S.}~\bibnamefont{Wong}},
  \bibinfo{journal}{Nucl. Phys.} \textbf{\bibinfo{volume}{A530}},
  \bibinfo{pages}{555} (\bibinfo{year}{1991}).

\bibitem[{\citenamefont{Melnitchouk et~al.}(1994)\citenamefont{Melnitchouk,
  Schreiber, and Thomas}}]{Melnitchouk:1994rv}
\bibinfo{author}{\bibfnamefont{W.}~\bibnamefont{Melnitchouk}},
  \bibinfo{author}{\bibfnamefont{A.~W.} \bibnamefont{Schreiber}},
  \bibnamefont{and} \bibinfo{author}{\bibfnamefont{A.~W.}
  \bibnamefont{Thomas}}, \bibinfo{journal}{Phys. Lett.}
  \textbf{\bibinfo{volume}{B335}}, \bibinfo{pages}{11} (\bibinfo{year}{1994}),
  \eprint{nucl-th/9407007}.

\bibitem[{\citenamefont{Braun and Tokarev}(1994)}]{Braun:1993nh}
\bibinfo{author}{\bibfnamefont{M.}~\bibnamefont{Braun}} \bibnamefont{and}
  \bibinfo{author}{\bibfnamefont{M.}~\bibnamefont{Tokarev}},
  \bibinfo{journal}{Phys. Lett.} \textbf{\bibinfo{volume}{B320}},
  \bibinfo{pages}{381} (\bibinfo{year}{1994}).

\bibitem[{\citenamefont{Burov and Molochkov}(1998)}]{Burov:1998kz}
\bibinfo{author}{\bibfnamefont{V.}~\bibnamefont{Burov}} \bibnamefont{and}
  \bibinfo{author}{\bibfnamefont{A.}~\bibnamefont{Molochkov}},
  \bibinfo{journal}{Nucl. Phys.} \textbf{\bibinfo{volume}{A637}},
  \bibinfo{pages}{31} (\bibinfo{year}{1998}).

\bibitem[{\citenamefont{Kulagin and
  Petti}(2006{\natexlab{b}})}]{Kulagin:2004ie}
\bibinfo{author}{\bibfnamefont{S.~A.} \bibnamefont{Kulagin}} \bibnamefont{and}
  \bibinfo{author}{\bibfnamefont{R.}~\bibnamefont{Petti}},
  \bibinfo{journal}{Nucl. Phys.} \textbf{\bibinfo{volume}{A765}},
  \bibinfo{pages}{126} (\bibinfo{year}{2006}{\natexlab{b}}),
  \eprint{hep-ph/0412425}.

\bibitem[{\citenamefont{Arrington
  et~al.}(2012{\natexlab{c}})\citenamefont{Arrington, Rubin, and
  Melnitchouk}}]{Arrington:2011qt}
\bibinfo{author}{\bibfnamefont{J.}~\bibnamefont{Arrington}},
  \bibinfo{author}{\bibfnamefont{J.}~\bibnamefont{Rubin}}, \bibnamefont{and}
  \bibinfo{author}{\bibfnamefont{W.}~\bibnamefont{Melnitchouk}},
  \bibinfo{journal}{Phys. Rev. Lett.} \textbf{\bibinfo{volume}{108}},
  \bibinfo{pages}{252001} (\bibinfo{year}{2012}{\natexlab{c}}),
  \eprint{1110.3362}.

\bibitem[{\citenamefont{Arrington et~al.}(2009)\citenamefont{Arrington,
  Coester, Holt, and Lee}}]{Arrington:2008zh}
\bibinfo{author}{\bibfnamefont{J.}~\bibnamefont{Arrington}},
  \bibinfo{author}{\bibfnamefont{F.}~\bibnamefont{Coester}},
  \bibinfo{author}{\bibfnamefont{R.}~\bibnamefont{Holt}}, \bibnamefont{and}
  \bibinfo{author}{\bibfnamefont{T.-S.} \bibnamefont{Lee}},
  \bibinfo{journal}{J. Phys.} \textbf{\bibinfo{volume}{G36}},
  \bibinfo{pages}{025005} (\bibinfo{year}{2009}), \eprint{0805.3116}.

\bibitem[{\citenamefont{Melnitchouk and Thomas}(1996)}]{Melnitchouk:1995fc}
\bibinfo{author}{\bibfnamefont{W.}~\bibnamefont{Melnitchouk}} \bibnamefont{and}
  \bibinfo{author}{\bibfnamefont{A.~W.} \bibnamefont{Thomas}},
  \bibinfo{journal}{Phys. Lett.} \textbf{\bibinfo{volume}{B377}},
  \bibinfo{pages}{11} (\bibinfo{year}{1996}), \eprint{nucl-th/9602038}.

\bibitem[{\citenamefont{Fenker et~al.}(2008)\citenamefont{Fenker, Burkert, Ent,
  Baillie, Evans et~al.}}]{Fenker:2008zz}
\bibinfo{author}{\bibfnamefont{H.~C.} \bibnamefont{Fenker}},
  \bibinfo{author}{\bibfnamefont{V.}~\bibnamefont{Burkert}},
  \bibinfo{author}{\bibfnamefont{R.}~\bibnamefont{Ent}},
  \bibinfo{author}{\bibfnamefont{N.}~\bibnamefont{Baillie}},
  \bibinfo{author}{\bibfnamefont{J.}~\bibnamefont{Evans}},
  \bibnamefont{et~al.}, \bibinfo{journal}{Nucl. Instrum. Meth.}
  \textbf{\bibinfo{volume}{A592}}, \bibinfo{pages}{273} (\bibinfo{year}{2008}).

\bibitem[{\citenamefont{Baillie et~al.}(2012)}]{Baillie:2011za}
\bibinfo{author}{\bibfnamefont{N.}~\bibnamefont{Baillie}} \bibnamefont{et~al.}
  (\bibinfo{collaboration}{CLAS Collaboration}), \bibinfo{journal}{Phys. Rev.
  Lett.} \textbf{\bibinfo{volume}{108}}, \bibinfo{pages}{199902}
  (\bibinfo{year}{2012}), \eprint{1110.2770}.

\bibitem[{\citenamefont{Tkachenko et~al.}(2014)}]{Tkachenko:2014byy}
\bibinfo{author}{\bibfnamefont{S.}~\bibnamefont{Tkachenko}}
  \bibnamefont{et~al.} (\bibinfo{collaboration}{CLAS Collaboration}),
  \bibinfo{journal}{Phys. Rev.} \textbf{\bibinfo{volume}{C89}},
  \bibinfo{pages}{045206} (\bibinfo{year}{2014}), \eprint{1402.2477}.

\bibitem[{\citenamefont{Accardi et~al.}(2011)\citenamefont{Accardi,
  Melnitchouk, Owens, Christy, Keppel et~al.}}]{Accardi:2011fa}
\bibinfo{author}{\bibfnamefont{A.}~\bibnamefont{Accardi}},
  \bibinfo{author}{\bibfnamefont{W.}~\bibnamefont{Melnitchouk}},
  \bibinfo{author}{\bibfnamefont{J.}~\bibnamefont{Owens}},
  \bibinfo{author}{\bibfnamefont{M.~E.} \bibnamefont{Christy}},
  \bibinfo{author}{\bibfnamefont{C.}~\bibnamefont{Keppel}},
  \bibnamefont{et~al.}, \bibinfo{journal}{Phys. Rev.}
  \textbf{\bibinfo{volume}{D84}}, \bibinfo{pages}{014008}
  (\bibinfo{year}{2011}), \eprint{1102.3686}.

\bibitem[{\citenamefont{Kalantarians et~al.}(2015)\citenamefont{Kalantarians,
  Christy, Ethier, and Melnitchouk}}]{Kalantarians:2015}
\bibinfo{author}{\bibfnamefont{N.}~\bibnamefont{Kalantarians}},
  \bibinfo{author}{\bibfnamefont{M.~E.} \bibnamefont{Christy}},
  \bibinfo{author}{\bibfnamefont{J.}~\bibnamefont{Ethier}}, \bibnamefont{and}
  \bibinfo{author}{\bibfnamefont{W.}~\bibnamefont{Melnitchouk}},
  \bibinfo{journal}{private communication, publication in preparation}
  (\bibinfo{year}{2015}).

\bibitem[{\citenamefont{Christy and Bosted}(2010)}]{Christy:2007ve}
\bibinfo{author}{\bibfnamefont{M.~E.} \bibnamefont{Christy}} \bibnamefont{and}
  \bibinfo{author}{\bibfnamefont{P.~E.} \bibnamefont{Bosted}},
  \bibinfo{journal}{Phys. Rev.} \textbf{\bibinfo{volume}{C81}},
  \bibinfo{pages}{055213} (\bibinfo{year}{2010}), \eprint{0712.3731}.

\bibitem[{\citenamefont{Bosted and Christy}(2008)}]{Bosted:2007xd}
\bibinfo{author}{\bibfnamefont{P.}~\bibnamefont{Bosted}} \bibnamefont{and}
  \bibinfo{author}{\bibfnamefont{M.~E.} \bibnamefont{Christy}},
  \bibinfo{journal}{Phys. Rev.} \textbf{\bibinfo{volume}{C77}},
  \bibinfo{pages}{065206} (\bibinfo{year}{2008}), \eprint{0711.0159}.

\bibitem[{\citenamefont{Arrington et~al.}(2006)\citenamefont{Arrington, Ent,
  Keppel, Mammei, and Niculescu}}]{Arrington:2003nt}
\bibinfo{author}{\bibfnamefont{J.}~\bibnamefont{Arrington}},
  \bibinfo{author}{\bibfnamefont{R.}~\bibnamefont{Ent}},
  \bibinfo{author}{\bibfnamefont{C.}~\bibnamefont{Keppel}},
  \bibinfo{author}{\bibfnamefont{J.}~\bibnamefont{Mammei}}, \bibnamefont{and}
  \bibinfo{author}{\bibfnamefont{I.}~\bibnamefont{Niculescu}},
  \bibinfo{journal}{Phys. Rev.} \textbf{\bibinfo{volume}{C73}},
  \bibinfo{pages}{035205} (\bibinfo{year}{2006}), \eprint{nucl-ex/0307012}.

\bibitem[{\citenamefont{Guzey et~al.}(2012)\citenamefont{Guzey, Zhu, Keppel,
  Christy, Gaskell et~al.}}]{Guzey:2012yk}
\bibinfo{author}{\bibfnamefont{V.}~\bibnamefont{Guzey}},
  \bibinfo{author}{\bibfnamefont{L.}~\bibnamefont{Zhu}},
  \bibinfo{author}{\bibfnamefont{C.~E.} \bibnamefont{Keppel}},
  \bibinfo{author}{\bibfnamefont{M.~E.} \bibnamefont{Christy}},
  \bibinfo{author}{\bibfnamefont{D.}~\bibnamefont{Gaskell}},
  \bibnamefont{et~al.}, \bibinfo{journal}{Phys. Rev.}
  \textbf{\bibinfo{volume}{C86}}, \bibinfo{pages}{045201}
  (\bibinfo{year}{2012}), \eprint{1207.0131}.

\bibitem[{\citenamefont{Owens et~al.}(2013)\citenamefont{Owens, Accardi, and
  Melnitchouk}}]{Owens:2012bv}
\bibinfo{author}{\bibfnamefont{J.}~\bibnamefont{Owens}},
  \bibinfo{author}{\bibfnamefont{A.}~\bibnamefont{Accardi}}, \bibnamefont{and}
  \bibinfo{author}{\bibfnamefont{W.}~\bibnamefont{Melnitchouk}},
  \bibinfo{journal}{Phys. Rev.} \textbf{\bibinfo{volume}{D87}},
  \bibinfo{pages}{094012} (\bibinfo{year}{2013}), \eprint{1212.1702}.

\bibitem[{\citenamefont{Frankfurt and Strikman}(1988)}]{Frankfurt:1988nt}
\bibinfo{author}{\bibfnamefont{L.}~\bibnamefont{Frankfurt}} \bibnamefont{and}
  \bibinfo{author}{\bibfnamefont{M.}~\bibnamefont{Strikman}},
  \bibinfo{journal}{Phys. Rept.} \textbf{\bibinfo{volume}{160}},
  \bibinfo{pages}{235} (\bibinfo{year}{1988}).

\end{thebibliography}
\bibliographystyle{apsrev}

\end{document}